\documentclass[12pt]{iopart}

\usepackage{citesort}

\begin{document}
\nocite{*}

\title{Gauge-invariant formulation for the gravitational wave equations}

\author{Junpei Harada}

\address{Health Sciences University of Hokkaido, 1757 Kanazawa, Tobetsu-cho, Ishikari-gun, Hokkaido 061-0293, Japan}
\ead{jharada@hoku-iryo-u.ac.jp}
\vspace{10pt}
\begin{indented}
\item[]May 2024
\end{indented}

\begin{abstract}
A gauge-invariant formulation for the gravitational wave equations is presented. Using this approach, weak, plane wave solutions in a vacuum are derived in various theories. These include general relativity with two modes of polarization with helicity $\pm 2$, Yang’s  theory with three modes of polarization with helicity $\pm 2$ and 0, and so-called “general metric theories” with six modes of polarization with helicity $\pm 2$, $\pm 1$, and two 0's. To identify the polarizations of gravitational waves, it is explicitly demonstrated how the gauge-invariant approach reproduces the earlier results. 
\end{abstract}

\section{Introduction}

To identify the polarizations of gravitational waves, the conventional approach uses the wave equations expressed in terms of the metric perturbation $h_{\mu\nu}$, along with some convenient gauge-fixing condition. This is the standard approach adopted in  textbooks (e.g.,~\cite{Weinberg:1972kfs,Wald:1984rg,Carroll:2004st}). However, gravitational waves are observables, suggesting it should be possible to express the gravitational wave equation in a gauge-invariant manner. Nevertheless, the author realized that this approach has been less frequently mentioned in the literature or textbooks.

Motivated by this observation, in this work, a gauge-invariant formulation for the gravitational wave equations is presented. Due to its gauge invariance, this formulation avoids the need for gauge fixing when dealing with the wave equations. Using this formulation, we derive weak, plane wave solutions in a vacuum for various theories: general relativity, Yang’s gravitational theory~\cite{Yang:1974kj}, and “general metric theories”~\cite{Eardley:1973br,Eardley:1973zuo} (explained later). Our results align with earlier results obtained through conventional methods. Thus, the results presented in this work are not new, while the formulation is original. The gauge-invariant approach and the conventional one are complementary, and both are useful.

In Sec.~\ref{sec:GR}, we discuss general relativity as an elementary example to formulate gauge-invariant wave equations. This is an elementary topic, yet the author could not find it in the standard textbooks. In Sec.~\ref{sec:Yang}, Yang’s theory~\cite{Yang:1974kj} is chosen to provide the simplest example for describing a polarization mode distinct from those in general relativity. In Sec.~\ref{sec:gmt}, more general theories are considered. It is commonly stated in the literature that there are six possible modes of polarization (two tensors, two vectors, and two scalars) in “general metric theories” for gravity~\cite{Eardley:1973br,Eardley:1973zuo}. We clarify the exact meaning of “general metric theories” in this context, and then derive six modes of polarization solely from the gauge-invariant wave equations. Section~\ref{sec:conclusion} provides a conclusion.

\section{Gravitational wave equation in general relativity\label{sec:GR}}

\subsection{Gauge-invariant wave equation}

In electromagnetism, the wave equations are given by
\numparts
\begin{eqnarray}
	 \opensquare A_\mu &=& 0, \label{eq:emwave_Lorenz}\\
	\partial_\mu A^\mu &=& 0 \label{eq:Lorenz}
 \end{eqnarray}
 \endnumparts
where $\opensquare=\partial^\mu\partial_\mu$ and $A_\mu$ is a vector potential. Here, (\ref{eq:emwave_Lorenz}) represents Maxwell's equations in a vacuum, and (\ref{eq:Lorenz}) is the gauge fixing condition proposed by Ludvig Lorenz. The equation~(\ref{eq:emwave_Lorenz}) holds only for $A_\mu$ in the Lorenz gauge. 

The gauge-invariant equations for describing the electromagnetic waves can be derived from the identities:
\begin{eqnarray}
	\partial_\rho F_{\mu\nu} + \partial_\mu F_{\nu\rho} + \partial_\nu F_{\rho\mu} = 0
	\label{eq:Fid}
\end{eqnarray}
where $F_{\mu\nu}=\partial_\mu A_\nu - \partial_\nu A_\mu$. Applying $\partial^\rho\equiv\eta^{\rho\mu}\partial_\mu$ to~(\ref{eq:Fid}) and using Maxwell's equations ($\partial_\mu F^{\mu\nu}=0$), we obtain
\begin{eqnarray}
	\opensquare F_{\mu\nu} = 0.\label{eq:emwave_inv}
\end{eqnarray}
Unlike~(\ref{eq:emwave_Lorenz}), this wave equation is gauge invariant because $F_{\mu\nu}$ is an Abelian field strength and thus gauge invariant. 

In the case of gravity, when gravity is weak, we can suppose that the metric is close to the flat metric $\eta_{\mu\nu}$:
\begin{eqnarray}
	g_{\mu\nu} = \eta_{\mu\nu} + h_{\mu\nu}
	\label{eq:metric}
\end{eqnarray}
 where $|h_{\mu\nu}| \ll 1$. To first order in $h$, the Ricci tensor is 
\begin{eqnarray}
 	R_{\mu\nu} = \frac{1}{2}\left(\partial_\mu\partial_\rho h^\rho{}_\nu + \partial_\nu\partial_\rho h^\rho{}_\mu - \partial_\mu\partial_\nu h^\lambda{}_\lambda - \opensquare h_{\mu\nu}\right).\quad
	\label{eq:Ricci_1st}
\end{eqnarray}
The Einstein equation $R_{\mu\nu}=0$ reduces to 
 \numparts
\begin{eqnarray}
	 \opensquare h_{\mu\nu} &=& 0, \label{eq:gw_Hilbert}\\
	\partial_\mu h^\mu{}_\nu &=& \frac{1}{2}\partial_\nu h^\lambda{}_\lambda \label{eq:Hilbert}
 \end{eqnarray}
 \endnumparts
where~(\ref{eq:Hilbert}) is known as the harmonic coordinate condition or the Hilbert gauge condition (sometimes referred to as the Lorenz gauge condition, but historically, “Hilbert” may be more appropriate). The wave equation~(\ref{eq:gw_Hilbert}) holds only for $h_{\mu\nu}$ satisfying~(\ref{eq:Hilbert}).

Equations~(\ref{eq:gw_Hilbert}) and~(\ref{eq:Hilbert}) correspond to~(\ref{eq:emwave_Lorenz}) and~(\ref{eq:Lorenz}), respectively. Our aim now is to find a gravitational analogue of $\opensquare F_{\mu\nu}=0$. 
 
Similar to electromagnetism, the gauge-invariant equations can be derived from the Bianchi identities:
 \begin{eqnarray}
	\partial_\lambda R_{\mu\nu\rho\sigma} + \partial_\rho R_{\mu\nu\sigma\lambda} + \partial_\sigma R_{\mu\nu\lambda\rho} = 0
	\label{eq:Bianchi}
\end{eqnarray}
where terms of ${\cal O}(h^2)$ are ignored. Throughout this paper, the covariant derivative $\nabla_\mu$ will be replaced with the partial derivative $\partial_\mu$, as the differences between the two derivatives are of higher order in $h$.

Applying $\partial^\lambda$ to equation~(\ref{eq:Bianchi}), we obtain
\begin{eqnarray}
	&&\opensquare R_{\mu\nu\rho\sigma} + \partial_\rho \partial^\lambda R_{\mu\nu\sigma\lambda} + \partial_\sigma \partial^\lambda R_{\mu\nu\lambda\rho} = 0.
	\label{eq:Bianchi2}
\end{eqnarray}
Using the identities $\partial^\lambda R_{\mu\nu\lambda\rho}=\partial_\mu R_{\nu\rho} - \partial_\nu R_{\mu\rho} + {\cal O}(h^2)$, equation (\ref{eq:Bianchi2}) reads:
\begin{eqnarray}
	&\opensquare R_{\mu\nu\rho\sigma} 
	- \partial_\mu\partial_\rho R_{\nu\sigma} + \partial_\nu\partial_\rho R_{\mu\sigma}
	 	+ \partial_\mu\partial_\sigma R_{\nu\rho} 
		- \partial_\nu\partial_\sigma R_{\mu\rho} = 0.
		\label{eq:Bianchi3}
\end{eqnarray}
When $R_{\mu\nu}=0$ holds, equation~(\ref{eq:Bianchi3}) reduces to 
\begin{eqnarray}
	\opensquare C_{\mu\nu\rho\sigma} = 0
	\label{eq:gw_GR}
\end{eqnarray}
where $C_{\mu\nu\rho\sigma}$ is the Weyl tensor. This is the wave equation in general relativity, expressed in terms of the curvature tensor rather than the metric perturbation $h_{\mu\nu}$.

To first order in $h$, it is straightforward to confirm that the Riemann tensor (and consequently the Weyl tensor) is invariant under the gauge transformations:
\numparts
\begin{eqnarray}
x^\mu \rightarrow x^{\prime \mu} &=& x^\mu + \varepsilon^\mu(x),\label{eq:gaugetrans_a}\\
h_{\mu\nu} \rightarrow h_{\mu\nu}^\prime &=& h_{\mu\nu} - \partial_\mu \varepsilon_\nu(x) - \partial_\nu \varepsilon_\mu(x)
\label{eq:gaugetrans_b}
\end{eqnarray}
\endnumparts
where $|h_{\mu\nu}^\prime| \ll 1$. Therefore, equation~(\ref{eq:gw_GR}) is gauge invariant. In this sense, the wave equation~(\ref{eq:gw_GR}) is a gravitational analogue of $\opensquare F_{\mu\nu}=0$. 

The Weyl tensor $C_{\mu\nu\rho\sigma}$ vanishes identically in two or three dimensions. Therefore, according to the wave equation~(\ref{eq:gw_GR}), gravitational waves do not propagate in two or three dimensions. However, in four or more dimensions, gravitational waves do propagate. We will now examine the plane wave solutions in four dimensions.

\subsection{Polarizations of gravitational wave}

The general solution of equation~(\ref{eq:gw_GR}) is a linear superposition of solutions of the form
\begin{eqnarray}
	C_{\mu\nu\rho\sigma} = e_{\mu\nu\rho\sigma} \exp (ik_\lambda x^\lambda) 
	+ {\rm c.c.}
	\label{eq:sol_Weyl_wave}
\end{eqnarray}
where “c.c.” denotes the complex conjugate, and $e_{\mu\nu\rho\sigma}$ is the polarization tensor. This satisfies equation~(\ref{eq:gw_GR}) if
\begin{eqnarray}
	k_\lambda k^\lambda = 0.
\end{eqnarray}

The Weyl tensor $C_{\mu\nu\rho\sigma}$ consists of ten independent components in four dimensions, and so does the polarization tensor $e_{\mu\nu\rho\sigma}$. Without loss of generality, we can choose, for example, the ten independent components as:
\begin{eqnarray}
	e_{0101}, e_{0102}, e_{0123}, e_{0131}, e_{0231},
	e_{1223}, e_{1231}, e_{1212}, e_{1201}, e_{1202}.
	\label{eq:ten_polarization}
\end{eqnarray}
To first order in $h$, these components are invariant under the transformations~(\ref{eq:gaugetrans_a}) and~(\ref{eq:gaugetrans_b}). 
When $R_{\mu\nu}=0$, the number of independent components of $e_{\mu\nu\rho\sigma}$ should reduce from {\it ten} to {\it  two}, corresponding to two physical degrees of freedom in general relativity. This will be demonstrated below. 
 
When $R_{\mu\nu}=0$, the Bianchi identities~(\ref{eq:Bianchi}) read
 \begin{eqnarray}
	\partial_\lambda C_{\mu\nu\rho\sigma} + \partial_\rho C_{\mu\nu\sigma\lambda} + \partial_\sigma C_{\mu\nu\lambda\rho} = 0.
	\label{eq:Bianchi_Weyl}
\end{eqnarray}
Contracting the indices of this equation yields
 \begin{eqnarray}
	\partial^\mu C_{\mu\nu\rho\sigma} = 0.
	\label{eq:Bianchi_Weyl_transverse}
\end{eqnarray}
The solution~(\ref{eq:sol_Weyl_wave}) satisfies~(\ref{eq:Bianchi_Weyl}) if 
 \begin{eqnarray}
	k_\lambda e_{\mu\nu\rho\sigma} + k_\rho e_{\mu\nu\sigma\lambda} + k_\sigma e_{\mu\nu\lambda\rho} = 0
	\label{eq:Bianchi_wavevector}
\end{eqnarray}
and satisfies~(\ref{eq:Bianchi_Weyl_transverse}) if
 \begin{eqnarray}
	k^\mu e_{\mu\nu\rho\sigma} = 0.
\end{eqnarray}
Thus, the polarization tensor $e_{\mu\nu\rho\sigma}$ is orthogonal to the wave vector $k^\mu$, indicating that the wave is transverse. 

For instance, consider a wave traveling in the $+z$ direction, with the wave vector
\begin{eqnarray}
	k^\mu = (k, 0, 0, k), \quad k>0.
	\label{eq:wave_vector}
\end{eqnarray}
In this case, the conditions~(\ref{eq:Bianchi_wavevector}) allow us to express the $e_{\mu\nu\rho\sigma}$ listed in~(\ref{eq:ten_polarization}) in terms of $e_{0101}$ and $e_{0102}$:
\begin{eqnarray}
	&e_{0131} = -e_{0101}, \nonumber \\
	& e_{0123} = -e_{0231} = e_{0102},\nonumber \\
	&e_{1223} = e_{1231} =  e_{1212} = e_{1201} = e_{1202}= 0.
	\label{eq:gr_independent_polarization}
\end{eqnarray}
Hence, only $e_{0101}$ and $e_{0102}$ are independent.

To find the helicity of gravitational waves, we consider a rotation about the $z$-axis:
\begin{eqnarray}
	e_{\mu\nu\rho\sigma}^\prime 
	= \Lambda_\mu{}^\alpha \Lambda_\nu{}^\beta \Lambda_\rho{}^\gamma \Lambda_\sigma{}^\delta
	e_{\alpha\beta\gamma\delta},
	\label{eq:Lorentz_trans}
\end{eqnarray}
where
\begin{eqnarray}
	&\Lambda_1{}^1=\Lambda_2{}^2=\cos\theta,\quad
	 \Lambda_1{}^2=-\Lambda_2{}^1=\sin\theta, \nonumber \\
	&\Lambda_0{}^0=\Lambda_3{}^3=1, \quad
	{\rm other}  \ \Lambda_{\mu}{}^\nu = 0.\label{eq:rotation}
\end{eqnarray}
From equations~(\ref{eq:Lorentz_trans}) and~(\ref{eq:rotation}), we find that 
\begin{eqnarray}
	e_{0101}^\prime \mp i e_{0102}^\prime = \exp (\pm 2 i \theta) (e_{0101} \mp i e_{0102}),
\end{eqnarray}
representing that the waves have helicity $\pm 2$. 

In summary, we have shown that the wave equation in general relativity is expressed by~(\ref{eq:gw_GR}), and the plane wave solution~(\ref{eq:sol_Weyl_wave}), along with~(\ref{eq:gr_independent_polarization}), represents two transverse modes of polarization with helicity $\pm 2$. To first order in $h$, the polarization tensor $e_{\mu\nu\rho\sigma}$ is invariant under the gauge transformation. Thus, this formulation is free from gauge fixing, as specified in~(\ref{eq:Hilbert}). 

\section{Scalar wave in Yang's theory\label{sec:Yang}}

In general relativity, the Ricci tensor vanishes in a vacuum, and gravitational waves are described by the Weyl tensor $C_{\mu\nu\rho\sigma}$, as shown in~(\ref{eq:gw_GR}). However, in certain gravitational theories, the Ricci tensor may not necessarily vanish even in a vacuum. In such theories, additional modes of polarization can arise. In this section, we take Yang’s gravitational theory as an example of such cases.

Yang's field equations~\cite{Yang:1974kj} in a vacuum are given by
\begin{eqnarray}
	\nabla_\mu R_{\nu\rho} = \nabla_\nu R_{\mu\rho}.
	\label{eq:Yang_EOM1}
\end{eqnarray}
From the Bianchi identities, this is equivalent to 
\begin{eqnarray}
	\nabla_\lambda R^\lambda{}_{\rho\mu\nu} = 0.
	\label{eq:Yang_EOM2}
\end{eqnarray}
Furthermore, this can be expressed as
\begin{eqnarray}
	2\nabla_\lambda C^\lambda{}_{\rho\mu\nu} + \frac{1}{6}(g_{\rho\mu}\partial_\nu - g_{\rho\nu}\partial_\mu)R = 0.
	\label{eq:Yang_EOM3}
\end{eqnarray}
We can readily observe that all vacuum solutions of the Einstein equations ($R_{\mu\nu}=0$ or $R_{\mu\nu}=\Lambda g_{\mu\nu}$) satisfy~(\ref{eq:Yang_EOM1}), and all vacuum solutions of Nordstr\"om's theory ($C_{\mu\nu\rho\sigma}=0, R=0$) satisfy~(\ref{eq:Yang_EOM3}).

When Yang's equation holds, Eq.~(\ref{eq:Bianchi2}) simplifies to 
\begin{eqnarray}
	\opensquare R_{\mu\nu\rho\sigma} = 0
	\label{eq:gw_Yang}
\end{eqnarray}
where $R_{\mu\nu\rho\sigma}$ satisfies~(\ref{eq:Yang_EOM2}). This is the wave equation in Yang's theory. To first order in $h$, this is gauge invariant. 

\subsection{Riemann tensor}
We consider weak, plane wave solutions to Eq.~(\ref{eq:gw_Yang}). The general solution of (\ref{eq:gw_Yang}) is as follows:
\begin{eqnarray}
	R_{\mu\nu\rho\sigma} = e_{\mu\nu\rho\sigma} \exp (ik_\lambda x^\lambda) 
	+ {\rm c.c}.
	\label{eq:sol_Yang_wave_riemann}
\end{eqnarray}
This satisfies~(\ref{eq:gw_Yang}) if $k_\lambda k^\lambda = 0$, 
and satisfies~(\ref{eq:Yang_EOM2}) if
\begin{eqnarray}
	k^\mu e_{\mu\nu\rho\sigma} = 0.
	\label{eq:Yang_Riemann}
\end{eqnarray}
Thus, the wave is null and transverse. The solution~(\ref{eq:sol_Yang_wave_riemann}) also satisfies the Bianchi identities~(\ref{eq:Bianchi}), and then we have
\begin{eqnarray}
	k_\lambda e_{\mu\nu\rho\sigma} + k_\rho e_{\mu\nu\sigma\lambda} + k_\sigma e_{\mu\nu\lambda\rho} = 0.
	\label{eq:Yang_Bianchi}
\end{eqnarray}

The Riemann tensor $R_{\mu\nu\rho\sigma}$ consists of twenty independent components in four dimensions, as does the polarization tensor $e_{\mu\nu\rho\sigma}$. However, when Yang's equation holds, the number of independent components of $e_{\mu\nu\rho\sigma}$ reduce from {\it twenty} to {\it three}, corresponding to three physical degrees of freedom. Hence, another polarization, distinct from the two tensor modes, appears as follows.

Similarly to Sec.~\ref{sec:GR}, consider a wave propagating in the $+z$ direction, with the wave vector $k^\mu=(k,0,0,k)$. In this case, Eqs.~(\ref{eq:Yang_Riemann}) and~(\ref{eq:Yang_Bianchi}) enable us to express twenty independent $e_{\mu\nu\rho\sigma}$ in terms of $e_{0101}$, $e_{0102}$, and $e_{0202}$:
\begin{eqnarray}
	&&e_{3131} = -e_{0131} = e_{0101},\nonumber\\
	&&e_{0123} = -e_{0231} = -e_{2331} = e_{0102},\nonumber\\
	&&e_{0223} = e_{2323} = e_{0202},\quad
	{\rm other} \ e_{\mu\nu\rho\sigma} = 0.
\end{eqnarray}
Hence, $e_{0101}$, $e_{0102}$, and $e_{0202}$ are independent, corresponding to three modes of polarization.

To find the helicity of the waves, we consider a rotation about the $z$-axis. From~(\ref{eq:Lorentz_trans}) and~(\ref{eq:rotation}), we find that 
\begin{eqnarray}
	&&\frac{e_{0101}^\prime-e_{0202}^\prime}{2} \mp i e_{0102}^\prime 
	= \exp (\pm 2i\theta) 
	\left(\frac{e_{0101}-e_{0202}}{2} \mp i e_{0102}\right)
	\label{eq:tensor_mode}
\end{eqnarray}
representing two tensor modes with helicity $\pm 2$, and 
\begin{eqnarray}
	e_{0101}^\prime+e_{0202}^\prime = e_{0101}+e_{0202}
	\label{eq:scalar_mode_b}
\end{eqnarray}
representing a scalar mode with helicity~0. Thus, in Yang's theory, there is a scalar mode in addition to two tensor modes.
These three modes are null and transverse. 

\subsection{Ricci tensor\label{sec:Ricci}}
Another convenient approach to derive the scalar mode is available as follows. 

When $R_{\mu\nu}=0$, equation~(\ref{eq:gw_Yang}) for the Riemann tensor simplifies to equation~(\ref{eq:gw_GR}) for the Weyl tensor. In this case, the solutions are identical to those of general relativity. Therefore, to identify additional polarization modes beyond those in general relativity, it is sufficient to consider the case when $R_{\mu\nu}\not=0$.

Contracting the indices in equation~(\ref{eq:gw_Yang}) yields
\begin{eqnarray}
	\opensquare R_{\mu\nu} = 0
	\label{eq:gw_Yang2}
\end{eqnarray}
where $R_{\mu\nu}$ satisfies~(\ref{eq:Yang_EOM1}).
The plane wave solution of~(\ref{eq:gw_Yang2}) is a linear superposition of solutions of the form
\begin{eqnarray}
	R_{\mu\nu} = e_{\mu\nu} \exp (ik_\lambda x^\lambda) 
	+ {\rm c.c}.
	\label{eq:sol_Yang_wave}
\end{eqnarray}
This satisfies~(\ref{eq:gw_Yang2}) if $k_\lambda k^\lambda = 0$ and satisfies~(\ref{eq:Yang_EOM1}) if
\begin{eqnarray}
	k_\mu e_{\nu\rho} = k_\nu e_{\mu\rho}.
	\label{eq:Yang_condition}
\end{eqnarray}

The Ricci tensor consists of ten independent components. This number should reduce from {\it ten} to {\it one}, corresponding a scalar mode as follows. Consider a wave propagating in the $+z$ direction with the wave vector $k^\mu=(k,0,0,k)$. The condition~(\ref{eq:Yang_condition}) gives:
\begin{eqnarray}
	&&e_{33} = - e_{03} = e_{00},\nonumber\\
	&&e_{01} = e_{02} = e_{11} = e_{22} = e_{23} = e_{31} = e_{12} = 0.
\end{eqnarray}
Thus, only $e_{00}$ is independent, and $e_{\mu\nu}$ is traceless. 

Under a rotation about the $z$-axis, $e_{\mu\nu}$ transforms as
\begin{eqnarray}
	e_{\mu\nu}^\prime = \Lambda_\mu{}^\alpha \Lambda_\nu{}^\beta e_{\alpha\beta}
\end{eqnarray}
where $\Lambda_\mu{}^\nu$ is given by~(\ref{eq:rotation}), and so we find that
\begin{eqnarray}
	e_{00}^\prime = e_{00}.
\end{eqnarray}
Hence, $e_{00}$ describes a scalar mode with helicity~0. 

In summary, we have shown that plane wave solutions in Yang's theory have three modes of polarization: two tensor modes with helicity~$\pm 2$, and a scalar mode with helicity~0. This scalar mode is described by the traceless Ricci tensor. These three modes are null and transverse.  

\section{Six polarizations in general metric theories\label{sec:gmt}}
In the literature, it is commonly stated that there are six modes of polarization (two tensors, two vectors, and two scalars) in 
“general metric theories” of gravitation. This conclusion was initially reached in Ref.~\cite{Eardley:1973br,Eardley:1973zuo} using the Newman--Penrose formalism~\cite{Newman:1961qr}. In this section, we begin by clarifying the meaning of “general metric theories” in this context. Then, we derive six modes of polarization solely from the gauge-invariant wave equations. 

\subsection{Riemann tensor}
We find from~(\ref{eq:Bianchi2}) that if the following relation
\begin{eqnarray}
	\partial_\rho \partial^\lambda R_{\mu\nu\sigma\lambda} + \partial_\sigma \partial^\lambda R_{\mu\nu\lambda\rho} = 0
	\label{eq:gmt_cond1}
\end{eqnarray}
is satisfied, then equation~(\ref{eq:Bianchi2}) reduces to 
\begin{eqnarray}
	\opensquare R_{\mu\nu\rho\sigma} = 0.
	\label{eq:gw_gmt}
\end{eqnarray}
Gravitational theories that {\it necessarily} satisfy Eq.~(\ref{eq:gmt_cond1}), such as general relativity or Yang’s theory, are referred to as “general metric theories” in this context. 

The plane wave solution of~(\ref{eq:gw_gmt}) takes the form
\begin{eqnarray}
	R_{\mu\nu\rho\sigma} = e_{\mu\nu\rho\sigma} \exp (ik_\lambda x^\lambda) 
	+ {\rm c.c.}
	\label{eq:sol_gmt_riemann}
\end{eqnarray}
This satisfies~(\ref{eq:gw_gmt}) if
\begin{eqnarray}
	k_\lambda k^\lambda = 0
	\label{eq:null_wave_vector}
\end{eqnarray}
and satisfies~(\ref{eq:gmt_cond1}) if
\begin{eqnarray}
	k_\rho k^\lambda e_{\mu\nu\sigma\lambda} + k_\sigma k^\lambda e_{\mu\nu\lambda\rho} = 0.
	\label{eq:gmt_cond1_k}
\end{eqnarray}
From the Bianchi identities~(\ref{eq:Bianchi}), we also have
\begin{eqnarray}
	k_\lambda e_{\mu\nu\rho\sigma} + k_\rho e_{\mu\nu\sigma\lambda} + k_\sigma e_{\mu\nu\lambda\rho} = 0.
	\label{eq:gmt_Bianchi}
\end{eqnarray}

These conditions~(\ref{eq:null_wave_vector}), (\ref{eq:gmt_cond1_k}), and (\ref{eq:gmt_Bianchi}) are not independent: When~(\ref{eq:gw_gmt}) holds, equation~(\ref{eq:gmt_cond1}) is automatically satisfied via the identities~(\ref{eq:Bianchi2}). Therefore, for a null wave vector $k^\mu$ satisfying~(\ref{eq:null_wave_vector}), considering~(\ref{eq:gmt_Bianchi}) alone is sufficient. It reduces the number of independent components of $e_{\mu\nu\rho\sigma}$ from {\it twenty} to {\it six} as follows.

Consider a wave propagating in the $+z$ direction with the wave vector $k^\mu=(k,0,0,k)$. In this case,  equation~(\ref{eq:gmt_Bianchi}) allows us to express twenty independent components of $e_{\mu\nu\rho\sigma}$ in terms of six $e_{\mu\nu\rho\sigma}$:
\begin{eqnarray}
	&&e_{3131} = - e_{0131} = e_{0101}, \nonumber\\
	&&e_{2323} =   e_{0223} = e_{0202}, \nonumber\\
	&&e_{0331} = -e_{0301},\nonumber\\
	&&e_{0323} = e_{0203},\nonumber\\
	&&e_{0123} =   -e_{0231} = -e_{2331} = e_{0102},\nonumber\\	
	&&e_{0112} = e_{0212} = e_{0312} = e_{2312} = e_{3112} = e_{1212} = 0.\quad
\end{eqnarray}
Six independent components, $e_{0101}$, $e_{0102}$, $e_{0202}$, $e_{0203}$, $e_{0301}$, and $e_{0303}$, correspond to six modes of polarization. 

Under a rotation about the $z$-axis, the transformations of three transverse modes, $e_{0101}$, $e_{0202}$, and $e_{0102}$, are given in equations~(\ref{eq:tensor_mode}) and~(\ref{eq:scalar_mode_b}) in Sec.~\ref{sec:Yang}. They represent two tensor modes with helicity $\pm 2$, and a scalar mode with helicity~0.

The other three non-transverse components, $e_{0203}$, $e_{0301}$, and $e_{0303}$, transform as
\begin{eqnarray}
	e_{0203}^\prime \pm i e_{0301}^\prime &=& \exp (\pm i\theta)(e_{0203} \pm i e_{0301})
\end{eqnarray}
representing two vector modes with helicity~$\pm 1$, and 
\begin{eqnarray}
	e_{0303}^\prime &=& e_{0303}	
\end{eqnarray}
representing a scalar mode with helicity~0. Thus, there are six modes of polarization (three of them are transverse, ant the others are non-transverse).

\subsection{Ricci tensor}
As in Sec.~\ref{sec:Ricci}, another convenient approach is to consider the wave equation associated with the Ricci tensor. This method is useful for clarifying additional modes beyond two tensor modes in general relativity.

Contracting the indices in~(\ref{eq:gw_gmt}) yields
\begin{eqnarray}
	\opensquare R_{\mu\nu} = 0
	\label{eq:gmt_wave_Ricci}
\end{eqnarray}
where $R_{\mu\nu}$ satisfies
\begin{eqnarray}
 \partial_\mu\partial_\rho R_{\nu\sigma} - \partial_\nu\partial_\rho R_{\mu\sigma}-\partial_\mu\partial_\sigma R_{\nu\rho} 
  +\partial_\nu\partial_\sigma R_{\mu\rho} = 0.\quad
  \label{eq:gmt_cond2}
\end{eqnarray}
To first order in $h$, this relation is identical to~(\ref{eq:gmt_cond1}).

The weak, plane wave solution of~(\ref{eq:gmt_wave_Ricci}) takes the form
\begin{eqnarray}
	R_{\mu\nu} = e_{\mu\nu} \exp (ik_\lambda x^\lambda) 
	+ {\rm c.c.}
	\label{eq:sol_gmt_Ricci}
\end{eqnarray}
This satisfies~(\ref{eq:gmt_wave_Ricci}) if $k_\lambda k^\lambda = 0$, and satisfies~(\ref{eq:gmt_cond2}) if
\begin{eqnarray}
 k_\mu k_\rho e_{\nu\sigma} - k_\nu k_\rho e_{\mu\sigma} - k_\mu k_\sigma e_{\nu\rho} 
  +k_\nu k_\sigma e_{\mu\rho} = 0. \quad
  \label{eq:gmt_cond3}
\end{eqnarray}
Two vector modes and two scalar modes should be described by~(\ref{eq:sol_gmt_Ricci}). 
Thus, the number of independent $e_{\mu\nu}$ should reduce from {\it ten} to {\it four} as follows. 

Consider a wave propagating in the $+z$ direction with the wave vector $k^\mu=(k,0,0,k)$. Equation~(\ref{eq:gmt_cond3}) gives six relations:
\begin{eqnarray}
	&e_{01} = - e_{31}, \nonumber\\
	& e_{02} = - e_{23}, \nonumber\\
	&e_{03} = - \frac{1}{2}(e_{00}+e_{33}),\quad
	&e_{11} = e_{22} = e_{12} = 0.
\end{eqnarray}
Hence, $10-6=4$ components of $e_{\mu\nu}$ are independent, corresponding to four additional modes of polarization.

Under a rotation about $z$-axis, we find that 
\numparts
\begin{eqnarray}
	&&e_{23}^\prime \pm i e_{31}^\prime = \exp (\pm i\theta)(e_{23} \pm i e_{31}), \\
	&&e_{00}^\prime = e_{00}, \quad e_{33}^\prime = e_{33}
\end{eqnarray}
\endnumparts
representing two vector modes with helicity~$\pm 1$, and two scalar modes with helicity~0.

\section{Conclusion\label{sec:conclusion}}
The gauge-invariant formulation for the gravitational wave equations has been presented. Using this approach, weak, plane wave solutions have been derived to identify polarization modes in various theories. These theories include general relativity with two tensor modes of polarization including helicity~$\pm 2$, Yang’s theory with three polarization modes including helicity~$\pm 2$ and~0, and so-called “general metric theories” with six polarization modes, including helicity $\pm 2$, $\pm 1$, and two $0$'s. Our results agree with the earlier results. Thus, the results presented in this study are not new, while the formulation is an original work. The gauge-invariant formulation presented here complements the conventional one using metric perturbation $h_{\mu\nu}$, and both approaches are useful.

\section*{Acknowledgments}
This work was supported by JSPS KAKENHI Grant No. JP22K03599.

\section*{References}
\bibliography{refereces-GW}
\end{document}